# A Rational Framework to Estimate the Chiroptical Activity of [6]Helicene Derivatives

Mirko Vanzan[‡], Susanna Bertuletti[‡], Belen Bazan, Minze T. Rispens, Steven I. C. Wan, Michel Leeman, Willem L. Noorduin, Francesca Baletto*

**ABSTRACT:** Helicenes are a class of molecules potentially suitable in several technological applications with intrinsic structural chirality which makes them interesting scaffolds for chiroptical properties. As desirable is tuning chiroptical property by synthesis, we combine experimental optical characterization and ab-initio calculations to study how different substituents influence the optical properties of [6]helicene. We explore anchoring groups presenting a variety of sizes and chemical nature, finding that both electron withdrawing and donating groups redshift and dwindle the optical activity of the molecule. We suggest the observed dumping in transitions energy and intensity is connected to the strength of the perturbation induced by the substituent on the π-conjugation of the aromatic rings. Such observations demonstrate how helicenes' chiroptical properties can be fine-tuned *via* stereochemical control of the substituents and validates a simple yet effective computational setup to model the optics of those systems.

Helicenes are a class of organic compounds whose backbone consists of a series of ortho-fused phenyl rings.[1,2] This particular configuration forces the molecule to assume an helicoidal shape, deviating from the planar conformation usually adopted by aromatic compounds. Ever since their discovery in the early 1900s, the chance of studying such peculiar molecular structures has stimulated the organic chemistry community to find fast and effective methods to synthetize those compounds.[2–4] However, it is since the 1970s that it is possible to produce numerous and increasingly complex helicenes, due to the introduction of synthetic strategies such as Diels-Alder reactions, ring-closing metathesis, metal-catalyzed coupling and photo-induced cyclizations.[5–10]

Nowadays, helicene chemistry is extensively investigated as these molecular scaffolds may be exploited in a highly diversified panel of applications such as metal complexes, molecular magnetism, nanorobotics, renewable energies and molecular photonics.[1,11–14]

Helicenes show indeed unique opto-electronic properties, especially when interacting with polarized light, because of their intrinsic chirality. As the helicity of the backbone can be either left- or right-handed, these molecules can exist in two different enantiomeric forms, thus providing a benchmark for the investigation of inherent chirality and chiral photonics. To date, a theoretical model able to capture the molecular optical features starting from its structure (and vice versa) is still unavailable due to the variety and complexity helicenes can present. However, being able to finely control the chiroptical response of those systems is of fundamental interest and represents a key-step toward their efficient use in technologically relevant applications. From this perspective, ab-initio simulations based on Time Dependent Density Functional Theory (TDDFT) represent a valuable tool to predict and optimize helicenes' opto-electronic properties as they can reasonably reproduce the time-dependent optoelectronic response of molecular systems with moderate computational costs.[15] A first seminal work in this sense comes from Furche and co-workers, who were the able to calculate [n]helicenes electronic circular dichroism (ECD) spectra by means TDDFT in the early 2000s.[16] Since then, the interest in the *ab initio* treatment of these compounds has progressively grown, and currently a vast amount of ECD spectra and g-factor estimates are available in the literature.[17–20]

In this work we propose a link between helicenes' chiroptical properties (intended as absorption, ECD, and g-factor spectra) and their substituent sizes and chemical nature. To do that, we apply a hybrid approach combining experimental and computational characterizations on a panel of derivates of [6]helicene. We focus on 5-mono-substituted molecules, where the substituent group could be either an amino ($NH_2$), a tert-butyloxycarbonyl (NHBoc), a bromide (Br) or a formyl (CHO). From now on, [6]helicene and its derivates will be called according to the numbering code in Figure 1. Although small, this pool of functional groups has variety in size (*e.g.*, NHBoc is larger than $NH_2$), and in chemical nature, being the group either electron donor (as in the case of $NH_2$), or withdrawing (like CHO). Such a variety allows to correlate the nature of the anchoring group to the optical response of the whole system. Notice that the numbering code in Figure 1 lists the substituents from the most activating ($NH_2$) to the most deactivating (CHO) group from the point of view of electrophilic aromatic substitutions, as shown also by the Hammet constants $\sigma_p$ in Table 1. We focused on the compounds left-handed (*m*-) enantiomers, and the substituent is always bound to the same carbon atom, preventing eventual spurious comparisons connected to differences in the group's relative position. Moreover, the chosen functional groups are all achiral, thus preventing the molecular chiroptical responses to be affected by the presence of stereogenic centres.

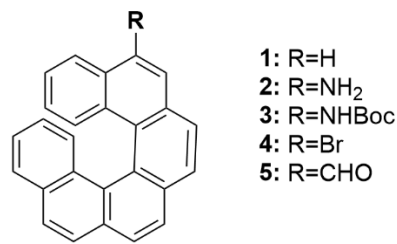

**Figure 1.** Schematic representation of the [6]-helicenes investigated in this work.

1: R=H
2: R=$NH_2$
3: R=NHBoc
4: R=Br
5: R=CHO

Apart from bare [6]helicene[21–24], the molecules selected for this study have been poorly investigated. To the best of our knowledge, there is only a single publication reporting an ECD spectra for **2**[25], while no results are available in the case of **3** and **5**. Regarding **4**, there are a few references showing its importance in helicene chemistry and its optical activity, but refer to different isomers.[26–28] This work provides not only a new theoretical framework to interpret the optical properties of functionalized helicenes, but also novel experimental and computational results on common, yet poorly studied, helicene compounds.

Experimental data are obtained on dichloromethane solutions of newly synthesized molecules. Details about synthesis and characterization are given in the Supporting Information (SI). Optical calculations are performed by means of real-time TDDFT.[29–32] As it is known that the solvent may influence the chiroptical activity of helicenes,[33,34] all calculations account for the presence of dichloromethane as solvent through an implicit solvent model.[35–37] To obtain a better agreement between theoretical and experimental bands, the theoretical absorption and ECD spectra have been systematically blue shifted by 60 nm, as commonly done in the literature for these kind of systems.[19,24,38,39] All the computational details are given in the SI.

The experimental and theoretical UV-Vis absorption spectra of the molecules are shown in Figure 2a. All compounds present similar optical features, with a main adsorption band falling within 300-375 nm, with the fine structure of the spectra depending on the specific molecular species. This is particularly visible from the experimental results (left side of Figure 2a). Compound **3** presents the most intense adsorption peaks, followed by **1**, **4**, **2** and **5**. The intense absorption of molecule **3** can be related to its size -significantly larger than the other compounds- and to the amide bond, reflecting in a larger absorption cross section and additional absorption in the 340-350 nm.

Concerning the position of the experimental (theoretical) main adsorption peaks, they fall at 316 (323) nm, 329 (330) nm, 332 (330) nm, 336 (339) nm and 336 (340) nm for molecules **1**, **4**, **3**, **2** and **5** respectively. This order reflects the magnitude of destabilization the group has on an aromatic ring. In fact, from an electronic point of view, Br (**4**), NHBoc (**3**), CHO (**5**) and NH$_2$ (**2**) increasingly perturb the π-conjugation of the aromatic ring system, weather because they have an electron donating or withdrawing nature. This is confirmed by an analysis of the partial atomic charges performed at the ground-state Density Functional Theory (DFT) level. We define Q$_H$ in the case of functionalized helicenes, as the sum of all C and H atoms partial Hirshfeld charges[40,41] not belonging to the substituent R, but constituting the helix backbone. As visible from Table 1, as the group provides more (positive or negative) charge on the aromatic rings, it accordingly affects the optical properties. Another common way to quantify π-perturbation in benzene-like aromatic systems is the substituents' Hammet parameters for para electrophilic additions (σ$_p$).[42] Such a number provides a qualitative measure of the effect a group has on the aromatic system and directly correlates with Q$_H$, as visible from Table 1. Together with the molecules DFT-optimized geometries, partial atomic Hirshfeld charges are available in the SI. To conclude the discussion on the absorption spectra, we notice that the most destabilizing groups (NH$_2$ and CHO) also give to the compounds a non-negligible adsorption even at longer wavelengths, with molecule **2** absorbing up to 500 nm. On the other hand, molecules **1** and **4** present a very similar adsorption fingerprint as Br weakly affects the electronic structure and thus the optical properties of the helicoid scaffold, as confirmed by its small Q$_H$ value (see Table 1). These observations represent a first direct proof of the influence the substituents nature can have on the optical properties of helicenes.

The considerations we have made on the shapes of the absorption spectra apply also to the TDDFT calculations (Figure 2a, right side) which, even though they do not reproduce the fine structure of the spectra, show the same relative trends. The theoretical *vs* experimental underestimation of the peaks intensity, in the order of 20-30% is connected to the chosen computational setup.[43]

**Table 1.** Substituents (R) Hammet constants (σ), helix backbone partial charges (Q$_H$), C-X excess charge difference (ΔQ$_{C-X}$) and Continuous Chirality Measure (CCM). Here C is the substituted carbon, and X is the R anchoring atom. Charges are given in units of [e], negative values indicate electron excess while positive values indicate electron depletion. σ$_p$ values are taken from ref.[44]

| Molecule | R | σ$_p$ | Q$_H$ | ΔQ$_{C-X}$ | CCM |
|---|---|---|---|---|---|
| 2 | NH$_2$ | -0.66 | -0.063 | 0.249 | 9.04 |
| 3 | NHBoc | -0.05 | -0.053 | 0.145 | 5.78 |
| 4 | Br | 0.23 | -0.005 | 0.010 | 9.97 |
| 5 | CHO | 0.42 | +0.091 | 0.110 | 9.18 |

The experimental and theoretical ECD spectra of the investigated compounds are shown in Figure 2b. Notice that the ECD profiles show negative values because we are studying the *m*-enantiomers of the compounds; *p*-enantiomers would present a mirrored trend on the positive values of Δε, where Δε is the difference in molar extinction coefficients for left- and right-handed circularly polarized light. Even in this case, all compounds present similar optical features, with a main band falling within a smaller wavelength window (310-350 nm), with the fine structure of the spectra depending on the specific molecular species. We notice that molecules **1**, **3** and **4** presents sharper Δε bands compared to **2** and **5**, as particularly visible from the experimental results (left side of Figure 2b). On the other hand, molecules **2** and **5** present a more sophisticated fine structures with some peaks present also around 300 nm. Compound **1** presents the most intense Δε peaks, followed by **4**, **3**, **1** and **2**. Generally speaking, the weaker the peaks the more redshifted are the wavelength of maximum Δε, as highlighted by the arrows in Figure 2b. The position of the experimental (theoretical) main peaks indeed, fall at 327 (327) nm, 330 (332) nm, 332 (333) nm, 336 (340) nm and 337 (339) nm for molecules **1**, **4**, **3**, **2** and **5** respectively. The only exception is represented by molecule **5** which, although shows a more intense peak, its maximum absorption wavelength is 1 nm longer than compound **2**. Interestingly, the similarities between UV-Vis spectra of compounds **1** and **4** are still held in the case of ECD, which spectra show very similar adsorption fingerprints. All these evidence find a physical rationale in the charge redistribution between the helicoidal scaffold and the substituent, as previously discussed. ECD indeed arises from chiral electronic transitions, which are directly influenced by the electronic density of the conjugated π-system. The higher the perturbation on the electronic structure of the aromatic rings is, the lower and less energetic the chiral response of the compound is. This confirms that the nature of the functional group affects also the chiroptical response of helicenes and in turn it demonstrates that it is possible to achieve fine tuning of the molecular optical properties by chemical modification of a small portion of the helix.

The considerations we have made on the shapes of ECD spectra apply also to the TDDFT calculations (Figure 2b, right side) which show the same relative trends. Even here, there is an underestimation of the peak's intensity as the theoretical vs experimental peaks are 10-15% less intense than the experimental ones. Notably, both calculated absorption and ECD spectra of bare [6]helicene (molecule 1) can be directly compared with previous calculations made with similar computational setups.[16,21,24]

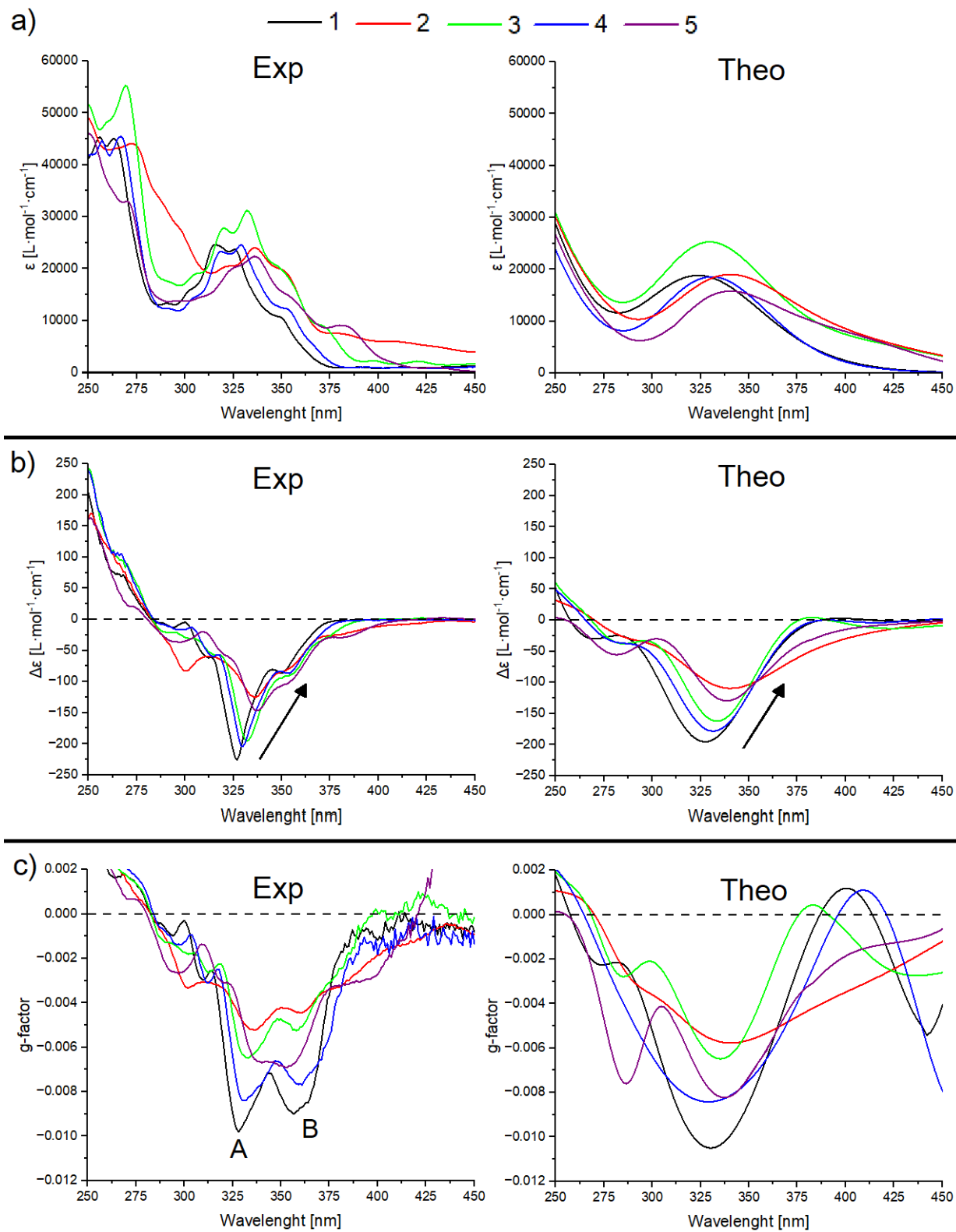

**Figure 2.** Collection of experimental (left side) and theoretical (right side) optical spectra and g-factor. a) UV-Vis absorption spectra, b) ECD spectra, inset arrows highlight the redshift of the main peak. c) g-factor. A and B label the two main peaks. Theoretical results are redshifted by 60 nm. Color code assigned according to the numbering code in Figure 1. One-by-one comparisons between experimental and theoretical results are available in the SI (Figure S1-S3).

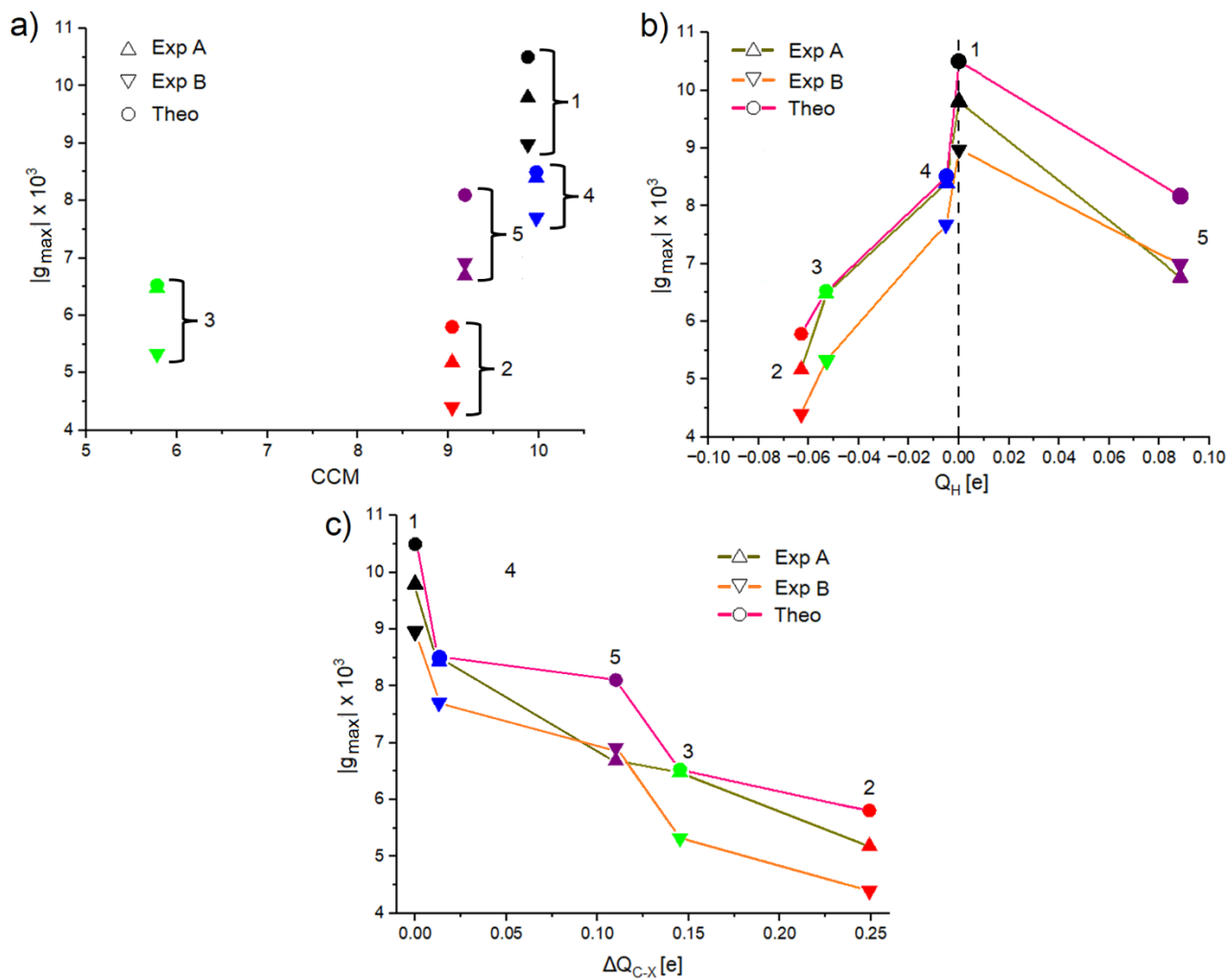

**Figure 3.** a), b) and c) show the correlation between the module of g-factor peaks intensity as a function of the CCM, $Q_H$ and $\Delta Q_{C-X}$. Exp A and Exp B refer to the experimental peaks of the g-factor (see Figure 2c). Symbols color code given accordingly to the legend of Figure 2, inset numbers label the molecule as per Figure 1.

The chirality of a compound is generally defined by its g-factor or dissymmetry factor. This is a dimensionless number quantifying the asymmetry in the absorption of left- and right-handed circularly polarized light and is defined as the ratio between $\Delta\varepsilon$ and $\varepsilon$ (molar extinction coefficient) as a function of the wavelength. Figure 2c shows the experimental and theoretical g-factors of the investigated compounds. The first thing that we notice is that the experimental g-factor profiles clearly show a double peak shape, with the one falling at shorter wavelength (peak A) being more intense than the one falling at longer wavelengths (peak B). The only exception is here represented by molecule **5**, where peak B is slightly more intense than peak A and the two-peak structure is barely visible.

The maximum absolute value of g-factors ($|g_{max}|$) is in the order of $10^{-2}$-$10^{-3}$, which is a typical value range for these kind of systems.[20] The g-factor$_{max}$ depends on the nature of the functional group, here following the trend **1**, **4**, **5**, **3**, **2**. Thus, bare [6]helicene the most chirally active among the compounds we study here. It is interesting to notice how the experimental $|g_{max}|$ fall in the same narrow wavelength windows, regardless of the molecules. In fact, considering peaks A (B), $|g_{max}|$ fall at 328 (357) nm, 330 (361) nm, 338 (353) nm, 333 (357) nm, 336 (360) nm for molecules **1**, **4**, **5**, **3** and **2** respectively. Both A and B peaks lie in a 10 nm-range, indicating a very stable peaks' profile.

The agreement between theoretical prediction and experiments holds in the case of calculated g-factor spectra (Figure 2c, right side) which, despite not catching the double peaks motif, closely follow the same experimental trend with the main peak falls in the 330-340 nm range. We predict an intensity 10-15% higher than the experimental values, due to our setup which underestimate $\varepsilon$ values[43] and hence an overestimation of the g-factor peaks. Such discrepancy in the intensity is more evident when ECD shows intense secondary peaks as in the case of molecule **5** (purple line in Figure 2c), which presents a strong, yet unreliable peak also around 280 nm.

The g-factor depends on the chiroptical properties of the molecules which depends on both their geometry and chemical features. To distinguish between these two effects, we analyse how $|g_{max}|$ correlates with those aspects. We quantify the geometrical chirality of our molecules through the Continuous Chirality Measure (CCM) parameter.[46,47] This number assigns a positive number to a structure, measuring the distance between the actual molecular configuration and the nearest achiral counterpart, with larger values indicating higher chirality degree. CCM values are summarized in Table 1 and the dependence of $|g_{max}|$ on CCM is shown in Figure 3a. From this plot is clear how, in general, higher degrees of geometrical chirality provide higher $|g_{max}|$, from both experimental and theoretical point of view. However, it is difficult to assess a direct proportionality between those quantities as there are exceptions to this rule. For example, despite showing larger CCM, molecules **2** and **4** provide weaker $|g_{max}|$ than **3** and **1** respectively. For the sake of completeness, consider the CCM of bare [6]helicene (**1**) is 9.88 and this value is compatible with previous estimations.[48] Interestingly, the compound with the largest achiral substituent (NHBoc, molecule **3**) is the one presenting the lowest CCM. Using achiral functional groups thus appears to be quenching the chirality of the whole structure, and this reduction would depend on the size of the group. Substituent size can therefore be a key-parameter when dealing with fine-tuning of the optical properties of helicenes. Taking a step forward, this also suggests that accounting for the same group size, functionalizing a helicene scaffold with a substituent presenting opposite chirality would reduce even more the systems' chiroptical response. To date we have no direct proves about this, however this may be the subject of future studies.

Figure 3b show the relationship between experimental and theoretical $|g_{max}|$ and $Q_H$, which resemble a volcano plot. Notice that here $Q_H$ for molecule **1** is here set to zero to provide a reference point for relative comparison, as the measure of $Q_H$ applies only in the case of functionalized helicenes. The maximum value of $|g_{max}|$ is obtained for [6]helicene, *i.e.*, where there are no heteroatoms perturbing the aromatic π-conjugation. Regardless of the nature of the functional group, we always notice a decrease in $|g_{max}|$ with the magnitude of this reduction depending on the amount of charge added or subtracted to the helicoid scaffold.

Our analysis suggests that electron donating groups tend to reduce both experimental and theoretical optical activities more than electron withdrawing substituents, as the slope of the two edges of the volcano plot are clearly different. However, it is necessary to underline that this is a qualitative trend and is by no means intended to be a quantitative measure of the substituents' effect. Indeed, we have only one case where $Q_H$ is positive, and this is insufficient to draw statistically meaningful conclusions. In any case, this qualitative observation indicates the C-X bond polarization (where X is the anchoring atom of the substituent group R) can affect the chiroptical properties of compounds. We therefore calculated $\Delta Q_{C-X}$, defined as the module of difference between partial C and X atomic charges. The higher this number the more polarized the bond is. Precise values are available in Table 1.

Figure 3c plots the correlation between experimental and theoretical $|g_{max}|$ and $\Delta Q_{C-X}$, showing that larger C-X bonds polarizations are connected to lower chiroptical activity, because of the larger perturbation induced on the aromatic π-conjugation. This means that a local electrostatic perturbation induced by the bond with the substituent can significantly change the physical properties of the helix. Even though there exist studies on the influence various functional groups have on helicenes' nonlinear optical features[38,49], to the best of our knowledge this is the first time a direct relation between helicenes' chiroptical properties and local bond polarization are assessed.

Although further studies are required, our work demonstrates how helicenes' chiroptical properties can be fine-tuned via stereochemical control of the substituents. Our results suggest that large achiral substituents decrease the geometrical chirality of the molecular structure and reduce its g-factor. Therefore, a proper design of the length and shape of the functional group may allow a precise control of the optical response of the systems, both in terms of absorption and ECD. Additionally, we find the chemical nature of the functional group heavily affects the optical response of the helicenes. In particular, the stronger the perturbation induced on the aromatic π-conjugation, both in terms of absolute charges localized on the rings and of carbon-substituent bond polarization, the more redshifted - the optically active transitions and the dwindle of ECD response and g-factor are.

## ASSOCIATED CONTENT

### Supporting Information

The Supporting Information is available free of charge on the ACS Publications website.

*Materials and methods: details on molecules synthesis, characterization and modelling; one-by-one comparison between experimental and computational results; DFT-optimized structures and partial atomic Hirshfeld charges of the investigated molecules.*

## AUTHOR INFORMATION

### *Corresponding Author

Francesca Baletto, Department of Physics, University of Milan, Via Celoria 16, 20133, Milan, Italy; ORCID: 0000-0003-1650-0010; Email: francesca.baletto@unimi.it

### Authors

Mirko Vanzan ‡, Department of Physics, University of Milan, Via Celoria 16, 20133, Milan, Italy; ORCID: 0000-0003-3521-8045

Susanna Bertuletti ‡, AMOLF, Science Park 104, 1098 XG, Amsterdam, The Netherlands; ORCID: 0000-0002-8231-3566

Belen Bazan, Symeres, 9747 AT Groningen, The Netherlands; ORCID: 0009-0008-9394-8883

Minze T. Rispens, Symeres, 9747 AT Groningen, The Netherlands; ORCID: 0009-0003-4656-2033

Steven I. C. Wan, Symeres, 9747 AT Groningen, The Netherlands; ORCID: 0000-0002-8998-9462

Michel Leeman, Symeres, 9747 AT Groningen, The Netherlands; ORCID: 0000-0002-8854-1084

Willem L. Noorduin, AMOLF, Science Park 104, 1098 XG, Amsterdam, The Netherlands; ORCID: 0000-0003-0028-2354

### Author Contributions

‡These authors contributed equally.

### Notes

The authors declare no competing financial interests.

## ACKNOWLEDGMENT

All authors acknowledge financial support from the European Commission under contract EIC Pathfinder CHIRALFORCE 101046961. M.V. acknowledges University of Milan for funding his postdoctoral fellowships "La bellezza degli aggregati: da nano a astro particelle". Computational resources provided by INDACO Platform, which is a project of High Performance Computing at the University of MILAN. We acknowledge also the CINECA award under the ISCRA initiative, for the availability of high performance computing resources and support.

Insert Table of Contents artwork here